\documentclass[twocolumn,amsmath,amssymb,aps,prl,superscriptaddress]{revtex4-1}

\usepackage{graphicx}
\usepackage{dcolumn}
\usepackage{bm}
\usepackage[colorlinks,urlcolor=blue,citecolor=blue,linkcolor=blue]{hyperref}
\usepackage{comment}
\usepackage{color}
\usepackage{physics}

\renewcommand{\k}{{\bf k}}

\newcommand{\q}{{\bf q}}

\newcommand{\0}{{\bf 0}}

\newcommand{\ab}{a_B}
\newcommand{\Nb}{N_B}

\newcommand{\eb}{\varepsilon_b}
\newcommand{\et}{\varepsilon_T}

\newcommand{\ek}{\epsilon_{\k}}

\newcommand{\nn}{\nonumber}
\newcommand{\beq}{\begin{equation}}
\newcommand{\eeq}{\end{equation}}

\newcommand{\sch}{Schr{\"o}dinger }

\newcommand{\new}[1]{{\color{black}#1}}

\begin{document}

\title{Quantum behavior of a heavy impurity strongly coupled to a Bose gas}

\author{Jesper Levinsen}
\affiliation{School of Physics and Astronomy, Monash University, Victoria 3800, Australia}
\affiliation{ARC Centre of Excellence in Future Low-Energy Electronics Technologies, Monash University, Victoria 3800, Australia}

\author{Luis A.~Pe{\~n}a~Ardila}
\affiliation{Institut f\"ur Theoretische Physik, Leibniz Universit\"at Hannover, Germany}

\author{Shuhei M.~Yoshida}
\affiliation{Biometrics Research Laboratories, NEC Corporation, Kanagawa 211-8666, Japan}

\author{Meera M.~Parish}
\affiliation{School of Physics and Astronomy, Monash University, Victoria 3800, Australia}
\affiliation{ARC Centre of Excellence in Future Low-Energy Electronics Technologies, Monash University, Victoria 3800, Australia}

\date{\today}

\begin{abstract} We investigate the problem of an infinitely heavy impurity interacting with a dilute Bose gas at zero temperature. When the impurity-boson interactions are short ranged, we show that boson-boson interactions induce a quantum blockade effect, where a single boson can effectively block or screen the impurity potential. Since this behavior depends on the quantum granular nature of the Bose gas, it cannot be captured within a standard classical-field description. Using a combination of exact quantum Monte Carlo methods and a truncated basis approach, we show how the quantum correlations between bosons lead to universal few-body bound states and a logarithmically slow dependence of the polaron ground-state energy on the boson-boson scattering length.  Moreover, we expose the link between the polaron energy and the spatial structure of the quantum correlations, spanning the infrared to ultraviolet physics.  \end{abstract}

\maketitle

The scenario of an infinitely heavy impurity in a quantum medium is a fundamental problem in physics, with relevance ranging from electron gases~\cite{MahanBook} to open quantum systems~\cite{Leggett1987}.  The behavior is well understood in the case of an ideal Fermi medium~\cite{Schmidt2018,Liu2020b} where the problem can be solved exactly. Here, Anderson famously demonstrated that any interaction with the impurity leads to the orthogonality catastrophe in the thermodynamic limit~\cite{Anderson1967}.  However, there is currently much debate over the nature of the ground state for a fixed impurity strongly coupled to a dilute Bose gas, which is of immediate importance to ongoing cold-atom experiments~\cite{Catani2012,Hu2016,Jorgensen2016,Camargo2018,SchmidtF2018,Yan2020,Skou2020}.

The bosonic problem --- termed the Bose polaron --- appears straightforward at first glance, since there is the possibility of describing the condensed ground state of the Bose gas as a classical field, e.g., in the form of a coherent state~\cite{Shchadilova2016,Loon2018,Drescher2019,Dzsotjan2020,Ardila2021}, or governed by an effective Gross-Pitaevskii equation~\cite{Drescher2020,Guenther2021,Massignan2021}.  Furthermore, when the Bose gas is non-interacting, the ground state corresponds to all bosons occupying the lowest single-particle state in the system, making it even simpler than the fermionic case~\cite{Fumi1955}.  However, this tendency of bosons to cluster also means that, in the absence of boson-boson interactions, the Bose polaron ground-state energy diverges when the impurity-boson interaction is attractive enough to support a bound state~\cite{Guenther2021,Drescher2021}.  Thus, it is an important and non-trivial question how this pathological behavior is cured by boson-boson interactions, and whether the details of the impurity-boson interaction play a key role.  This is of particular interest in the case of short-range resonant impurity-boson interactions, where the scattering length $a \to \pm \infty$ and there is the prospect of universal physics, independent of the microscopic details.

\begin{figure}
	\centering
	\includegraphics[width=0.83\linewidth]{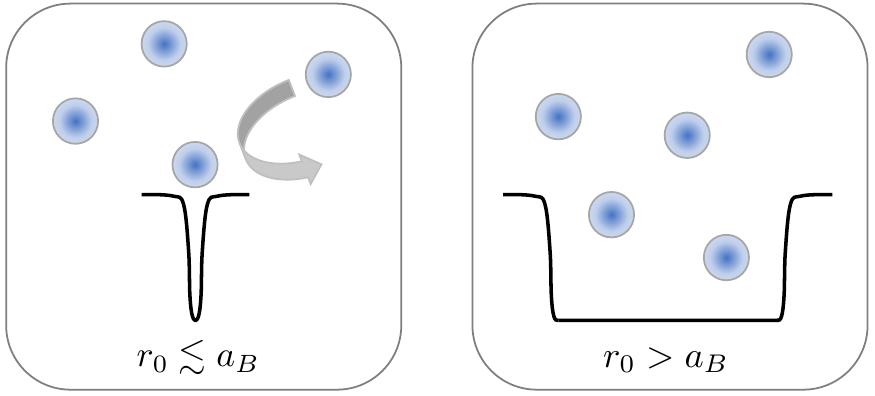}
	\caption{Bosons (circles) in the presence of an attractive impurity potential. If the range of the potential $r_0$ is comparable to or smaller than the boson-boson scattering length $\ab$, then a single boson can block the potential (left). Conversely, if $r_0 >\ab$, as for a Rydberg~\cite{Camargo2018} or ionic~\cite{Tomza2019} impurity, then many bosons can interact with the potential at once (right).}
	\label{fig:blocking}
\end{figure}

In this Letter, we show that in order to describe the ground state of the Bose polaron, it is crucial to go beyond classical-field descriptions and include the quantum ``granular'' nature of the Bose gas.  Specifically, once the boson-boson scattering length $\ab$ is comparable to or larger than the range $r_0$ of the attractive impurity-boson potential, a single boson from the gas can effectively screen or block the impurity potential, as illustrated in Fig.~\ref{fig:blocking}.  For a sufficiently attractive impurity-boson potential with $r_0 \to 0$, we find that this quantum blocking effect leads to universal few-body bound states involving the impurity, in agreement with Refs.~\cite{Shi2018,Yoshida2018PRA}.  Using exact quantum Monte Carlo (QMC) methods~\cite{Ardila2015,Ardila2016,Ardila2019}, we show that the polaron energy in the many-body limit exhibits a \textit{logarithmic} dependence on $\ab$ in the unitary regime $a \to \pm \infty$.  We further illustrate the importance of quantum correlations between bosons by showing that the QMC results for the polaron ground-state energy are well captured by a truncated basis variational approach~\cite{Levinsen2015,Yoshida2018PRX,Field2020} across a range of interactions.

\paragraph{Model.---} We consider the following Hamiltonian for a single infinitely heavy impurity in a Bose gas:
\begin{align}
    \hat H =\sum_\k \ek 
    b^\dag_\k 
    b_\k +
    \sum_{\k\k'\q}\frac{V(\q)}2
    b^\dag_{\k}
    b^\dag_{\k'}
    b_{\k'+\q}
    b_{\k-\q}+g\sum_{\k\k'}
    b^\dag_\k 
    b_{\k'}.
    \label{eq:Ham1}
\end{align}
The three terms correspond, respectively, to the kinetic energy of the bosons, the boson-boson interaction, and the boson-impurity interaction, where we have set the system volume and $\hbar$ to one.  In this model, a boson of mass $m$ and momentum $\k$ is created by the operator $
b^\dag_\k$, and we consider bosons with the quadratic dispersion $\ek=|\k|^2/2m\equiv k^2/2m$. Furthermore, we describe their interaction using the short-range potential $V(\q)$, which results in a low-energy boson-boson scattering length $a_B >0$. The interaction between the impurity and a boson is taken to be short-ranged and of strength $g$ up to a momentum cutoff $\Lambda$. The bare parameters $g$ and $\Lambda$ can be related to the physical impurity-boson scattering length $a$ via $ \frac m{2\pi a}=\frac1g+\sum_\k^\Lambda \frac1\ek$.  In the following, we take the zero-range limit $r_0 \to 0$, which requires $\Lambda \to \infty$. For the QMC calculations, we solve the problem in real space, using a Bethe-Peierls boundary condition for the impurity-boson interactions, and taking the boson-boson potential to be a hard-sphere potential, where the diameter of the sphere coincides with the s-wave scattering length $\ab$ (see Supplemental Material~\cite{supmat}).

\begin{figure} \centering
  \includegraphics[width=0.87\linewidth]{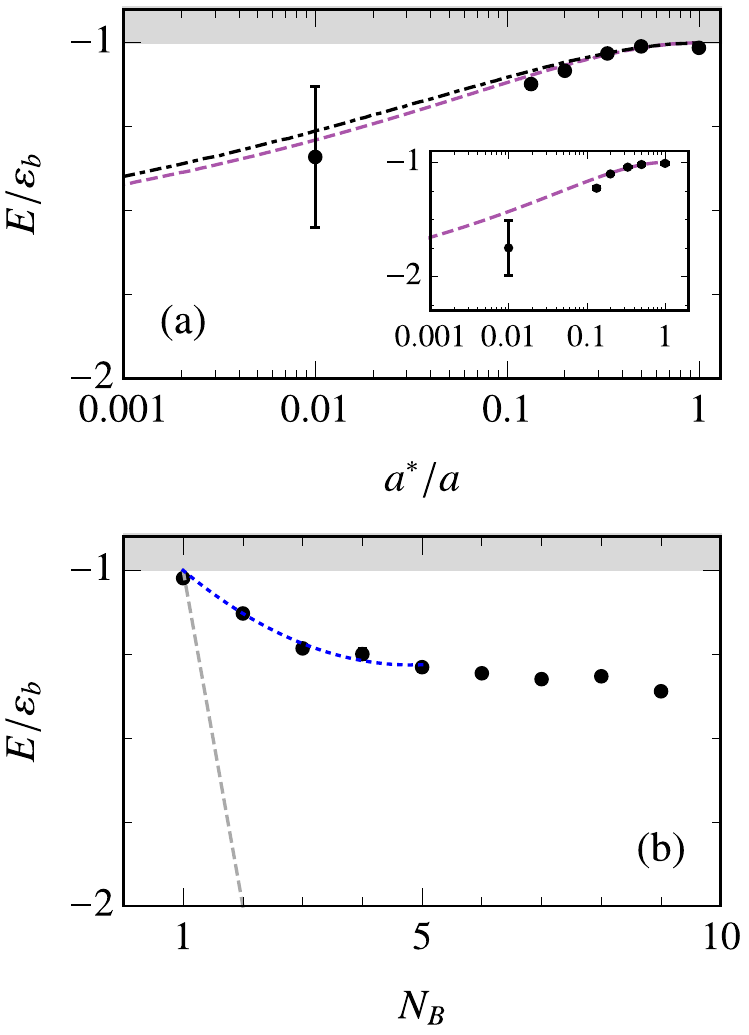} \caption{(a)
    Trimer energy as a function of inverse scattering length obtained
    from QMC (black circles), bosons with attractive contact
    interaction (black dot dashed line), and the Anderson model
    (purple dashed line). The inset compares the tetramer energy in
    the QMC with those of the Anderson model. (b) Few-body energy at
    $a/\ab=75$ as a function of boson number calculated within the QMC
    (black circles). We also show the energy of uncorrelated bosons,
    $E=-\Nb\eb$ (gray dashed line), and that of interacting bosons in
    an effective potential that accounts for three-body correlations,
    Eq.~\eqref{eq:int-boson}, with $U = 0.04\eb=1.5\ab/ma^3$ (blue
    dotted line). Data for the Anderson model is taken from
    Ref.~\cite{Shi2018}.}
	\label{fig:fewbody}
\end{figure}

\paragraph{Few-body bound states.---} We first discuss the few-body physics of an infinitely heavy impurity interacting with $N_B$ identical bosons, where we assume that $a>0$ such that the impurity potential supports a bound state.  For $N_B=1$, we simply have the impurity-boson bound state with energy $-\eb= -1/2ma^2$, while $N_B = 2$ corresponds to the minimal number of bosons where boson-boson correlations can emerge. In Fig.~\ref{fig:fewbody}(a) we display the QMC results for the $N_B = 2$ energy for a range of $\ab$. We find that a trimer (2-boson) bound state only exists when the scattering length $a$ is above a critical value $a^* \simeq 10 \ab$ set by the boson repulsion. Moreover, the trimer energy remains close to $-\eb$ (i.e., the result for $N_B=1$) for the plotted range of $\ab/a$ spanning several orders of magnitude, and it only slowly approaches the result for uncorrelated bosons, $-2\eb$, as we take $a_B/a \to 0$. A similar behavior is observed for $N_B = 3$, since we see that the tetramer (3-boson) bound state also only exists when $a>a^*$, and the tetramer energy lies well above the uncorrelated result, $-3\eb$. Therefore, we conclude that boson repulsion dominates the few-body behavior.

Indeed, we find that we can reproduce these few-body states when the bosons only block each other at the impurity and are non-interacting otherwise. Such a scenario is achieved with a bosonic Anderson model~\cite{Shi2018,Yoshida2018PRA}, where the impurity-boson interaction features an open and closed channel like in a realistic cold-atom scattering process~\cite{Chin2010}. Here, the impurity is unavailable for interactions with other bosons once a boson enters the closed-channel state, thus mimicking the quantum blockade effect in Fig.~\ref{fig:blocking}.  We previously solved the $N_B = 2$ problem exactly analytically for this model and we obtained the critical scattering length $a^*=3.1426|r_{\rm eff}|$, where $r_{\rm eff}$ is the (negative) effective range of the impurity-boson interactions~\cite{Shi2018,Yoshida2018PRA}. Moreover, we found that $a^*$ corresponded to a multibody resonance beyond which all $N_B >1$ bound states cease to exist. We display the results of this two-channel model in Fig.~\ref{fig:fewbody}(a) and find good agreement with the QMC data. This demonstrates two points: the few-body energies universally depend on the ratio $a^*/a$, and the behavior is determined by quantum blocking at the impurity.

Such few-body universality also extends to models with \textit{zero-range} boson-boson interactions. In this case, a finite positive $\ab$ requires an underlying attractive potential $V(\q)$, which features Efimov physics as well as deeply bound dimers~\cite{supmat}. Thus, the relevant few-body states with effective boson-boson repulsion are actually metastable excited states. Nonetheless, it is possible to solve for the energy of the metastable trimer state~\cite{supmat} and we see that it agrees well with the results of the other models in Fig.~\ref{fig:fewbody}(a). We also find the critical scattering length to be $a^*=20.0 a_B$, which differs slightly from that estimated from the QMC simulations for a hard-sphere potential, indicating that finite-range effects are relevant in the relationship between $a^*$ and boson repulsion.

Within QMC, we can extend our results to even larger $N_B$ complexes.  Fixing $a^*/a <1$, we observe in Fig.~\ref{fig:fewbody}(b) that the energy strongly deviates from the uncorrelated result $E=-N_B\eb$ (dashed gray line) and appears to saturate to a finite value with increasing $N_B$.  Moreover, this does not match the energy of interacting bosons in a potential, $E = -N_B\eb + U N_B (N_B-1)/2$, for \textit{any} interaction energy $U$.  We expect this behavior to also hold for a non-zero range $r_0$ as long as we satisfy the blocking condition $r_0 \lesssim \ab$, illustrated in Fig.~\ref{fig:blocking}. This condition is equivalent to requiring that the boson interaction energy, $\sim \ab/m r_0^3$, exceeds the depth of the potential, $\sim 1/mr_0^2$, assuming that the potential is close to resonance and using the fact that bosons within the potential interact over a volume set by $r_0^3$~\footnote{Note that for the case of an ionic impurity, we instead have $r_0 \gg \ab$ and thus the polaron energy does follow the relationship $E = -N_B \eb$ for a range of $N_B$~\cite{Astrakharchik2020}.}.

We can understand the result of Fig.~\ref{fig:fewbody}(b) by considering instead $N_B-1$ bosons moving in the longer-ranged potential originating from the infinitely heavy dimer consisting of the impurity and a boson.  In this case, the range of the effective potential is $\sim a$ and the energy of interacting bosons is
\begin{align} \label{eq:int-boson}
    E = -\eb - (N_B -1) \et  + \frac{U}{2} (N_B-1) (N_B-2), 
\end{align}
where $\et$ is the trimer binding energy. In Fig.~\ref{fig:fewbody}(b) where $a \gg \ab$ and $\et \ll \eb$, we see that the small-$N_B$ behavior is well captured by Eq.~\eqref{eq:int-boson} using $U \sim \ab/ma^3$.  This illustrates the importance of three-body correlations as well as demonstrating the role of the potential range.

\paragraph{Many-body limit.---} 
We now turn to the behavior of an impurity in a Bose gas of finite density $n$. In the absence of the impurity and in the limit of vanishing boson-boson interactions, the ground state is a Bose-Einstein condensate (BEC): $\ket{\Phi} = 
e^{\sqrt{n} ( b_\0^\dag - b_\0)} \ket{0}$, where $\ket{0}$ is the vacuum state for bosons. Thus, we can replace operators $b_\0^\dag$ and $b_\0$ in the Hamiltonian \eqref{eq:Ham1} by $\sqrt{n}$. Introducing the impurity and turning on interactions, the polaron ground state can be written in the general form~\cite{Yoshida2018PRX}
\begin{align} \label{eq:wfn} 
   \ket{\Psi}  = & \Big( \alpha_0 + \sum_{\k\neq \0} \alpha_\k b^\dag_\k + \frac{1}{2}\sum_{\k_1, \k_2 \neq \0} \alpha_{\k_1\k_2} b^\dag_{\k_1} b^\dag_{\k_2} \ldots \Big) \! \ket{\Phi} , 
\end{align}
where the complex coefficients $\alpha_j$ are associated with different numbers of bosons excited out of the condensate, and $\alpha_{\k_1\k_2}=\alpha_{\k_2\k_1}$.  In principle, one could write the expansion in Eq.~\eqref{eq:wfn} in terms of Bogoliubov excitations rather than bare bosonic excitations~\cite{Li2014,Levinsen2015}. However, this only modifies the operators at low momenta $k < 4 \sqrt{\pi n \ab}$, and this is not expected to affect the leading order behavior of the polaron energy in the extremely dilute limit $n^{1/3} \ab \ll 1$~\cite{Yoshida2018PRX}.  It is also likely that the Bogoliubov approximation breaks down in the regime of strong impurity-boson interactions~\cite{Grusdt2017,Ichmoukhamedov2019}.

Applying the Hamiltonian~\eqref{eq:Ham1} to the state~\eqref{eq:wfn} and keeping only the leading order boson-boson interaction terms in the limit $n^{1/3} \ab \ll 1$, we obtain the ground-state polaron energy~\cite{supmat}:
\begin{align} \label{eq:energy}
    E = n \left[\frac{m}{2\pi a} + \sum_\k\left(\frac{1}{\ek + G_\k} - \frac{1}{\ek}\right) \right]^{-1} .
\end{align}
Crucially, we find that it depends on the repulsive correlations between bosons via the positive function 
\begin{align} \label{eq:Gk}
G_\k = g\sqrt{n}\left(\sum_{\k'} \alpha_{\k\k'} /\alpha_\k - \sum_{\k'} \alpha_{\k'}/\alpha_0 \right)  .
\end{align}
Note that the case of uncorrelated non-interacting bosons corresponds to $\alpha_{\k\k'} = \alpha_\k \alpha_{\k'}/\alpha_0$, 
which gives $G_\k = 0$, such that the polaron energy $E =2\pi n a/m$, in agreement with previous work~\cite{Guenther2021,Drescher2021}.  Thus, the presence of correlations is necessary to ensure that the ground-state energy remains finite in the unitarity limit $1/a \to 0$. 

\begin{figure}
	\centering
	\includegraphics[width=0.87\linewidth]{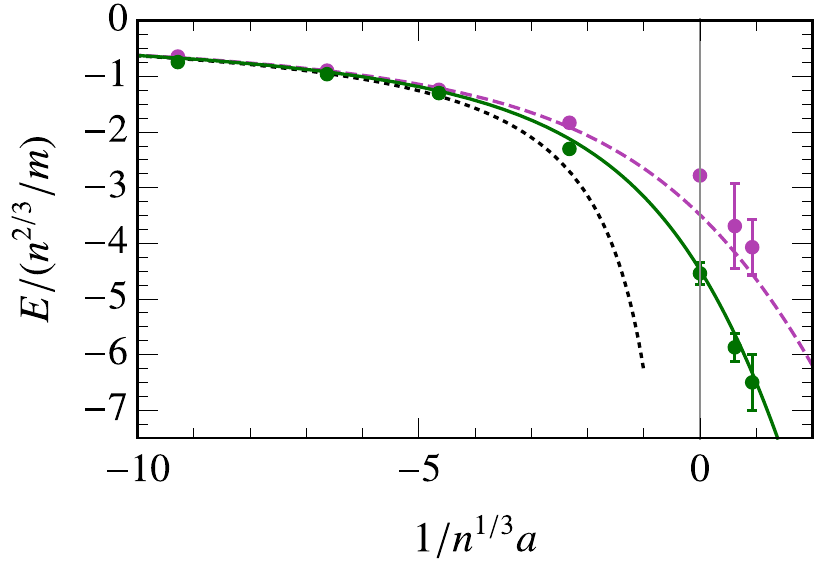}
	\caption{Ground-state energy of the infinitely heavy Bose polaron 
	as a function of inverse impurity-boson scattering length at fixed $n^{1/3}a^*=0.215$ (purple dashed) and $n^{1/3}a^*=0.00215$ (green solid). We show the results of the QMC (symbols) together with the results of the truncated basis 
	approach in the Anderson model with up to 3 excitations (lines). The mean-field result, $E =2\pi n a/m$, is depicted as a dotted line.}
	\label{fig:manybody}
\end{figure}

This behavior is confirmed in Fig.~\ref{fig:manybody}, where we display the polaron ground-state energy obtained using exact QMC methods for two different densities differing by two orders of magnitude.  For weak impurity-boson attraction $1/n^{1/3} a \ll -1$, we recover the mean-field uncorrelated result $E =2\pi n a/m$, which corresponds to the leading order dependence of Eq.~\eqref{eq:energy} on $a$. However, as anticipated, the energy becomes sensitive to boson-boson correlations as we increase the interactions towards unitarity.  This behavior is not just limited to zero-range impurity-boson interactions since the same result is obtained for a finite-range potential when $r_0 < \ab$~\cite{Ardila2015}. Note that this behavior goes beyond the few-body results discussed previously since the impurity-boson bound state is either absent (when $a<0$) or larger than the interparticle spacing ($n^{1/3}a \gtrsim 1$).

To further characterize the correlations, we also calculate the polaron energy using a variational approach~\cite{Yoshida2018PRX}, where we truncate the number of bosonic excitations in the polaron ground state in Eq.~\eqref{eq:wfn}~\cite{supmat}.  Here we again use the Anderson model to mimic the blockade effect at the impurity, and we use an effective range $r_{\rm eff} \simeq -3 \ab$.  This ensures that the value of the three-body parameter $a^*$ that quantifies the boson-boson repulsion matches the one from the QMC simulations. As shown in Fig.~\ref{fig:manybody}, we find that the truncated basis approach accurately reproduces the QMC results across a wide range of $n^{1/3} a^*$ (up to two orders of magnitude) when we include up to three excitations only.  This suggests that the boson-boson repulsion suppresses impurity-induced excitations of the condensate, and that this suppression is universal, \textit{i.e.}, independent of the microscopic origin of $a^*$. We stress that this is a highly quantum effect that cannot be captured by a classical mean-field description~\cite{Kalas2006}.

\begin{figure}
	\centering
	\includegraphics[width=0.87\linewidth]{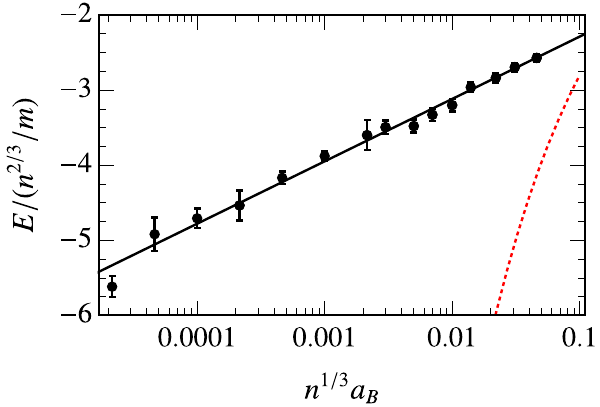}
	\caption{Bose polaron ground-state energy in the unitarity regime of impurity-boson interactions,  $1/a=0$. The QMC results (symbols) are consistent with a logarithmic dependence of the form $E^{\rm QMC} \simeq 
          0.36\ln(0.019n^{1/3}\ab)n^{2/3}/m$
	(solid line). The dashed red line is the prediction of the coherent state ansatz within the Bogoliubov approximation~\cite{Shchadilova2016}.}
	\label{fig:manybody-unitary}
\end{figure}

At unitarity $1/a = 0$, the polaron energy takes the universal form
\begin{align}
E = -f(n^{1/3}\ab) \, n^{2/3}/m,
\end{align}
where $f(x)$ is a dimensionless function. When $\ab \to 0$ at fixed density, we know that $E \to -\infty$, while in the zero-density limit $n \to 0$, we must have $E \to 0$ since there are no bound states. Thus, in the limit $n^{1/3} \ab \to 0$, we require $f(x) \to \infty$ slower than $\sim 1/x^2$. %
Indeed, our QMC results reveal a \textit{logarithmically} slow dependence $f(x) \sim -\ln(x)$, as shown in Fig.~\ref{fig:manybody-unitary}. This behavior is difficult to fully capture within the truncated basis approach~\cite{supmat} since it requires an increasingly larger number of boson excitations as $n^{1/3} \ab \to 0$.  On the other hand, if we use a coherent-state ansatz~\cite{Shchadilova2016} with an infinite number of excitations but only the approximate mean-field repulsion of the Bogoliubov Hamiltonian, then we have $f(x) =\sqrt{\pi/4x}$ which drastically overestimates the change in energy (see Fig.~\ref{fig:manybody-unitary}).  \new{The classical-field approach in Ref.~\cite{Massignan2021} also predicts a power-law behavior $f(x) \sim 1/x^{1/3}$, but this is only valid when $r_0 \gg a_B$, which is different from the regime considered here~\footnote{Note that Ref.~\cite{Massignan2021} claims that their results are valid even for short-ranged potentials $r_0 \sim a_B$, but this appears to neglect the condition that the number of bosons must be large in all modes of the classical field. Defining $n_l$ to be the local density at the impurity, we thus require $N_l = n_l r_0^3 \gg 1$ as well as the dilute gas condition $n_l a_B^3 \ll 1$, which gives the long-range requirement $r_0 \gg a_B$. See also Ref.~\cite{Chen2018}, which derived the condition $r_0 \gg a_B$ for the validity of the classical field approach in the zero-density limit.}.}

Indeed, the polaron energy is intimately connected to the spatial structure of the boson-boson correlations via the function $G_\k$ in Eq.~\eqref{eq:energy}, which can be viewed as an effective interaction potential between two excited bosons. In the infrared limit $k \to 0$, where the bosons are at large separation, we should recover the behavior of uncorrelated bosons. Here, we expect that the difference in energy between one and two excited bosons is their mean-field interaction with the condensate, $8\pi \ab n/m$.  This large-distance infrared behavior is correctly captured by the coherent state ansatz~\cite{Shchadilova2016}, which however fails at shorter length scales since it predicts a constant $G_\k=8\pi \ab n/m$ for all $k$ and $a$~\cite{supmat}.  In reality, we expect the blockade effect to dominate at short distances such that $\alpha_{\k\k'} \to 0$, and in this case one can show that $G_\k \to -E$ as $k \to \infty$~\cite{supmat}.  This short-distance ultraviolet behavior is captured by a ``Chevy-type'' ansatz with a single boson excitation~\cite{Chevy2006,Rath2013,Li2014}, but this ansatz does not describe the large-distance physics since it has $G_\k = -E$ at all momenta.  However, the momentum dependence of $G_\k$ can be well approximated within a truncated basis approach that includes more boson excitations~\cite{supmat}, as considered in this work.  In particular, our results indicate that quantum blocking at short distances dominates the behavior of the polaron energy while the infrared physics only provides a small correction.

\paragraph{Conclusion.---} To conclude, we have shown that the ground state of the Bose polaron exhibits strong quantum correlations between bosons when the impurity-boson potential is short-ranged. This is due to a quantum blockade effect at the position of the impurity, which gives rise to universal few-body bound states and a logarithmically slow dependence of the polaron energy on boson-boson interactions in the unitarity limit $1/a \to 0$.  Our results should be directly applicable to cold-atom experiments, where typically $r_0 \sim \ab$~\cite{Chin2010}, and they should also extend to a heavy but finite impurity mass $m_{I}$ since Efimov physics is exponentially suppressed as a function of $m_{I}/m$~\cite{Braaten2006}.  More generally, the Bose polaron scenario could provide a route to probing and engineering quantum correlations in other bosonic systems such as photons in microcavities~\cite{Munoz2019,Delteil2019}.

\acknowledgments
We gratefully acknowledge fruitful discussions with Nils-Eric Guenther, Victor Gurarie, Pietro Massignan, and Zheyu Shi. JL and MMP are supported through Australian Research Council Future Fellowships FT160100244 and FT200100619, respectively. JL and MMP also acknowledge support from the Australian Research Council Centre of Excellence in Future Low-Energy Electronics Technologies (CE170100039).

\bibliography{cold-atoms,other-refs}

\renewcommand{\theequation}{S\arabic{equation}}
\renewcommand{\thefigure}{S\arabic{figure}}
\renewcommand{\thetable}{S\arabic{table}}

\onecolumngrid


\setcounter{equation}{0}
\setcounter{figure}{0}
\setcounter{table}{0}

\clearpage

\section*{SUPPLEMENTAL MATERIAL:\\ ``Quantum behavior of a heavy impurity strongly coupled to a Bose gas''}
\setcounter{page}{1}
\begin{center}
Jesper~Levinsen,$^{1,2}$ 
Luis A.~Pe{\~n}a~Ardila,$^{3}$  
Shuhei M.~Yoshida,$^{4}$  and
Meera M.~Parish$^{1,2}$ \\
\emph{\small $^1$School of Physics and Astronomy, Monash University, Victoria 3800, Australia}\\
\emph{\small $^2$ARC Centre of Excellence in Future Low-Energy Electronics Technologies, Monash University, Victoria 3800, Australia}\\
\emph{\small $^3$ Institut f\"ur Theoretische Physik, Leibniz Universit\"at Hannover, Germany }\\
\emph{\small $^4$Biometrics Research Laboratories, NEC Corporation, Kanagawa 211-8666, Japan}

\end{center}

\section{Quantum Monte Carlo}

The Hamiltonian of a system of $N$ indistinguishable bosons and a single static impurity can be written as
\begin{align}
\mathcal{H}=-\frac{\hbar^{2}}{2m}\sum_{i=1}^{N}\nabla_{i}^{2}+\sum_{i<j}V_{BB}(r_{ij})+\sum_{i=1}^{N}V_{IB}(r_{i\alpha}),
\end{align}
where $m$ is the boson mass and $\alpha$ labels the impurity.  Here $V_{BB}$ and $V_{IB}$ are the boson-boson and impurity-boson potentials, which depend on the boson-boson $r_{ij}=\left|\mathbf{r}_{i}-\mathbf{r}_{j}\right|$ and impurity-boson $r_{i \alpha }=\left|\mathbf{r}_{\alpha}-\mathbf{r}_{i}\right|$ relative distances, respectively.

The  trial wave function of the system can be written as $ \psi(\mathbf{R})=\prod_{i<j}f_{B}(r_{ij})\prod_{i=1}^{N}f_{I}(r_{i\alpha})$, where  $f_{B}(r_{ij})$ and $f_{I}(r_{i \alpha})$ are the respective Jastrow boson-boson and impuirty-boson correlators. Here $\mathbf{R}=\left\{ \mathbf{r_{\alpha},r_{\text{1}},r_{\rm{2}},}\cdots,\mathbf{r_{\rm{N}}}\right\}$ represents the $3\times (N+1)$ vector containing the impurity and the bosons coordinates. The boson-boson Jastrow term is used as in Refs.~\cite{Ardila2015,Ardila2016,Ardila2019}. In the following section we discuss the specific choices for the potentials and Jastrow wave-functions.

\subsection{Boson-boson interactions}

We use a hard-sphere potential to model the repulsion between bosons. The potential reads,
\begin{align}
 V_{BB}(\mathbf{r})=\left\{ \begin{array}{c}
+\infty\quad r\leq \ab\\
0\quad  r>\ab
\end{array}\right.
\end{align}
 where the radius of the potential coincides with the boson-boson scattering length.  The Jastrow term for the atom-atom correlations is chosen by matching the solution of the two-body scattering problem,
 \begin{align}
f_{B}(r)=\left\{ \begin{array}{c}
0\\
\sin\left[q(r-a_{B})/r\right]
\end{array}\right.\begin{array}{c}
r\leq a_{B}\\
a_{B}<r<L/2
\end{array}
\end{align}
the vector $q$ by imposing continuity of the wave-function, in addition, in order to be consistent with periodic boundary conditions, the first derivative of the Jastrow function must be  zero at half of the box size $L=V^{1/3}$, where $V$ is the size of the simulation box.

\subsection{Impurity-boson interactions}

The impurity-boson Jastrow wave function is written as,
\begin{align}
 f_{I}(r)=\left\{ \begin{array}{c}
\psi^{BP}(r)\\
A+B\left(\exp\left[-\alpha r\right]+\exp\left[-\alpha(L-r)\right]\right)
\end{array}\right.\begin{array}{c}
r\leq\bar{R}\\
\bar{R}<r<L/2,
\end{array}
    \label{eq:QMC3}
\end{align}
where $\psi^{BP}=1-a/r$ in absence of bound state~\cite{Ardila2016}  or $\psi^{BP}=\exp\left(-kr\right)/r$ in terms of the energy of a two-body bound-state in vacuum $\varepsilon_{b}=\frac{k^{2}}{2m}=-\frac{1}{2ma^{2}}$. The wavefunction and its first derivative are continuous at the matching point $\bar{R}$,  where $\bar{R}=xL/2$, $x\in[0,1]$. The parameters $x$ and $\alpha$  are optimised using a variational Monte Carlo method. Note that, the Jastrow wave function $\psi^{BP}$ displays  a divergence at $r=0$ and hence we resort to the contact boundary Bethe-Peierls (BP) condition  $f_{I}(r)=\psi^{BP}$ which allow us to use a zero-range potential instead of using a finite-range one. This condition has been heavily used in QMC  for both Fermi~\cite{Pessoa15,Pessoa15-2} and Bose systems~\cite{Ardila2016,Ardila2020}.  

\section{Bosons with short-range attraction}

When the bosons in the medium have zero-ranged attractive interactions, the Hamiltonian in Eq.~\eqref{eq:Ham1} becomes
\begin{align}
    \hat H =\sum_\k \ek 
    b^\dag_\k 
    b_\k +
    \sum_{\k\k'\q}\frac{g_B}2
    b^\dag_{\k}
    b^\dag_{\k'}
    b_{\k'+\q}
    b_{\k-\q}+g\sum_{\k\k'}
    b^\dag_\k 
    b_{\k'},
    \label{eq:Ham1App}
\end{align}
with the constant $g_B <0$. In this case, the boson-boson scattering length $a_B$ is determined via
\begin{align}
    \frac m{4\pi a_B}=\frac1{g_B}+\sum_\k^{\Lambda_B} \frac1{2\ek},
\end{align}
with $\Lambda_B$ an ultraviolet cutoff. When $a_B>0$ there exists a two-boson bound state with energy $-1/ma_B^2$.

\subsection{Two- and three-body problems}
We start by making the following observation: The model in Eq.~\eqref{eq:Ham1App} features Efimov few-body physics for \textit{any} short-range attraction between bosons, rendering the model ultraviolet divergent. As a consequence, the infinitely heavy Bose polaron in an ideal Bose gas corresponds to a quantum critical point for Efimov physics, a point that has not previously been discussed in the literature. 

\begin{figure}
	\centering
	\includegraphics[width=.8\linewidth]{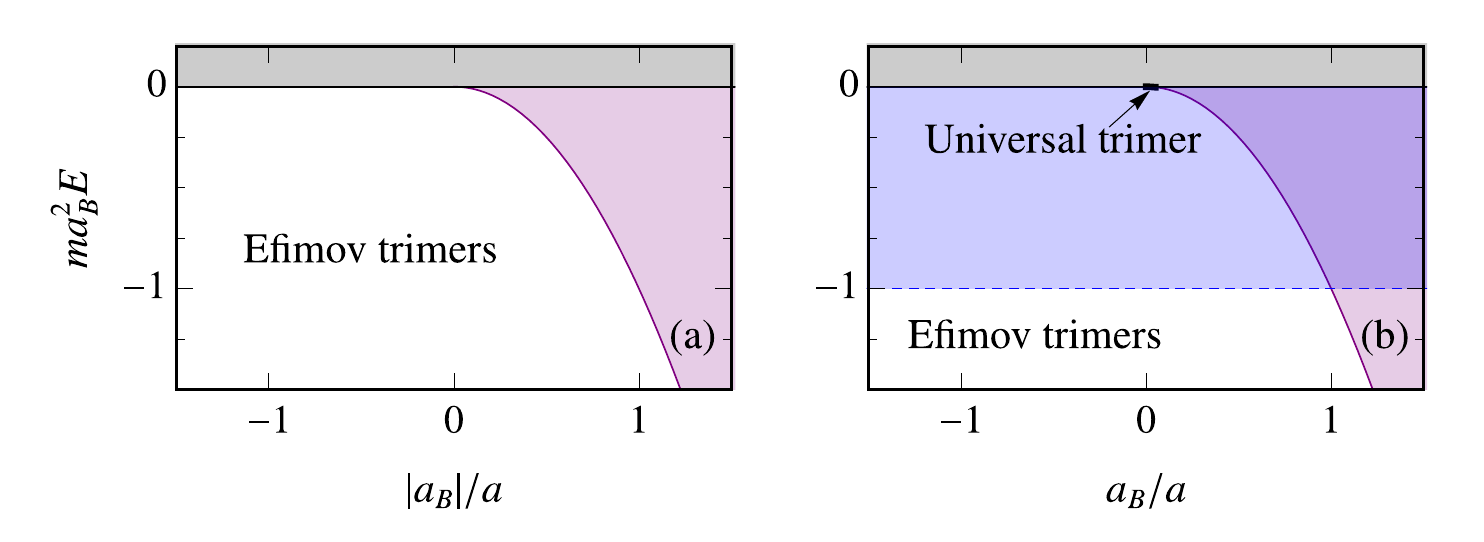}
	\caption{Few-body spectrum for an impurity and one or two bosons. The bosons interact with attractive zero-range interactions, resulting in either $a_B<0$ (a) or $a_B>0$ (b). We have a continuum above zero energy (gray area), above the impurity-boson bound state (purple area), and above the two-boson bound state in panel (b) (blue area). Below these continua, the three-body problem is unstable to the formation of Efimov trimers. In (b), we mark the existence of a small region close to unitarity-limited impurity-boson interactions where there is universal few-body physics.}
	\label{fig:schematic}
\end{figure}

The various regimes of the few-body spectrum for the case of $N_B=1$ and $N_B=2$ are illustrated in Fig.~\ref{fig:schematic}.  For $a_B<0$ in panel (a) we have a continuum above zero energy when $a<0$ and above the two-body impurity-boson bound state that exists when $a>0$. Below this continuum we have the existence of Efimov trimers and the spectrum is unbounded without an additional short-range cutoff. When $a_B>0$, panel (b), for $N_B=2$ we additionally have the boson-boson bound state with binding energy $1/ma_B^2$, leading to the horizontal line in the figure. Note that the Efimov effect can be quite significant even for the infinitely heavy impurity, and it does not necessarily involve huge scaling factors. For instance, in the case where the two scattering lengths are identical, $a=a_B$, and they greatly exceed any other length scale in the problem, the discrete scaling factor determining the Efimov spectrum is $\lambda_0=15.7$~\cite{Braaten2006,Naidon2017}, which implies that there exists an infinite tower of Efimov trimer states that can be mapped onto each other via the discrete transformation of the scattering lengths $a=a_B\to \lambda_0^l a=\lambda_0^la_B$ and energy $E\to E\lambda_0^{-2l}$ where $l$ is an integer. The scaling factor can thus be even smaller than the $\lambda_0= 22.7$ found in the case of three identical bosons with large scattering length~\cite{Efimov1970,Braaten2006}. On the other hand, the scaling factor is infinite when $a_B=0$~\cite{Efimov1973}.

What may appear puzzling in Fig.~\ref{fig:schematic} is the apparent absence of the  state where two bosons both occupy the boson-impurity bound state for $a>0$. Indeed, such a configuration trivially exists when the bosons are completely uncorrelated. However, we find this state only in a tiny subset of the already very small region marked ``universal trimer'' in panel (b). To see this, we solve explicitly the three-body problem consisting of the impurity and two bosons. Taking a general state
\begin{align}
    \ket{\Psi}=\frac12\sum_{\k_1\k_2}\alpha_{\k_1\k_2}b^\dag_{\k_1}b^\dag_{\k_2}\ket{0}
\end{align}
with $\alpha_{\k_1\k_2}=\alpha_{\k_2\k_1}$, we find that the \sch equation $(E-\hat H)\ket{\Psi}$ leads to
\begin{align}
    (E-\epsilon_{\k_1}-\epsilon_{\k_2})\alpha_{\k_1\k_2}=& \ g_B\sum_{\k_1'\k_2'}\alpha_{\k_1'\k_2'}\delta_{\k_1+\k_2-\k_1'-\k_2'} 
    +g\sum_{\k_1'}\alpha_{\k_1'\k_2}+g\sum_{\k_2'}\alpha_{\k_1\k_2'}.
\end{align}
Defining the quantities on the right hand side, $\eta_{\k}\equiv g_B\sum_{\k_1'\k_2'}\alpha_{\k_1'\k_2'}\delta_{\k-\k_1'-\k_2'}$ and $\chi_\k\equiv g\sum_{\k'}\alpha_{\k\k'}$, we obtain
\begin{subequations}
\label{eq:trimers}
\begin{align}
    {\cal T}^{-1}_{IB}(E\!-\!\ek)\chi_\k&=\sum_{\k'}\left(\frac{\eta_{\k'}}{E-\ek-\epsilon_{\k-\k'}}\!+\!\frac{\chi_{\k'}}{E-\ek-\epsilon_{\k'}}\right)\!, \label{eq:trimersa}\\
    {\cal T}^{-1}_{BB}(E\!-\!\ek/2)\eta_\k&=2\sum_{\k'}\frac{\chi_{\k'}}{E-\epsilon_{\k'}-\epsilon_{\k-\k'}}.\label{eq:trimersb}
\end{align}
\end{subequations}
Here we have defined the vacuum impurity-boson and boson-boson scattering $T$ matrices
\begin{subequations}
\begin{align}
    {\cal T}^{-1}_{IB}(E) & = \frac m{2\pi}(1/a-\sqrt{-2mE-i0}),
    \\ {\cal T}^{-1}_{BB}(E) &=\frac m{4\pi}(1/a_B-\sqrt{-mE-i0}),
\end{align}
\end{subequations}
respectively. 

Equation~\eqref{eq:trimers} represents a set of coupled integral equations that can, for instance, be solved as an eigenvalue equation for $\chi_\k$ with eigenvalue $1/a$ by replacing $\eta_\k$ in Eq.~\eqref{eq:trimersa} by its representation in Eq.~\eqref{eq:trimersb}.  In order to uniquely determine the spectrum of Efimov trimers, one must additionally supply a cutoff on the sums on the right-hand-side of the equation. However, our focus is on the universal trimer which exists for large values of the ratio $a/a_B$ when $a_B>0$, and which is independent of such a cutoff. This trimer is unusual, as it is sitting in a continuum of states corresponding to the relative motion of a two-boson bound state relative to the impurity, see Fig.~\ref{fig:schematic}(b). These continuum states sensitively depend on the choice of grid whereas the trimer state is independent of this choice. Consequently, the trimer state is easily distinguishable from the continuum states in our numerical solution of Eq.~\eqref{eq:trimers}, and we simply discard the continuum states from our consideration.

\section{Bose polaron ground state}

For the model in Eq.~\eqref{eq:Ham1} of the main text with generic repulsive boson-boson interactions, the Hamiltonian for the Bose polaron is
\begin{align} \label{eq:Ham-pol}
    \hat{H} = \sum_\k \ek b^\dag_\k b_\k  + g \sqrt{n} \sum_{\k\neq\0}(b^\dag_\k + b_{-\k}) + g \sum_{\k,\q\neq\0} b^\dag_\k b_\q + g n +  \sum_{\k\k'\q}\frac{V(\q)}2 b^\dag_{\k}b^\dag_{\k'} b_{\k'+\q} b_{\k-\q}.
\end{align}
Here we have replaced the operator $b_\0$ by $\sqrt{n}$, the square root of the condensate density, in the terms involving the impurity. Since we will focus on the scenario of an extremely dilute Bose gas near the resonance of impurity-boson interactions where $n^{1/3}\ab \ll 1$ while $n^{1/3} |a| \gtrsim 1$, we do not perform the Bogoliubov prescription on the last term of the Hamiltonian. Instead, we use a complete description of the boson-boson interactions that is needed to render the model ultraviolet convergent in this limit.

We obtain the polaron energy $E$ by applying the Hamiltonian to the general state in Eq.~\eqref{eq:wfn} of the main text,
\begin{align} \label{eq:wfn2}
   \ket{\Psi}  = & \Big( \alpha_0 + \sum_{\k\neq \0} \alpha_\k b^\dag_\k + \frac{1}{2}\sum_{\k_1,\k_2\neq \0} \alpha_{\k_1\k_2} b^\dag_{\k_1} b^\dag_{\k_2} + \frac{1}{3!}\sum_{\k_1,\k_2,\k_3 \neq \0} \alpha_{\k_1\k_2\k_3} b^\dag_{\k_1} b^\dag_{\k_2} b^\dag_{\k_3}  + \ldots \Big) \! \ket{\Phi}  ,
\end{align}
and then projecting onto different numbers of bosons excited out of the condensate. This yields the (infinite) set of coupled equations:  
\begin{subequations} \label{eq:pol}
\begin{align} \label{eq:pol1}
    E\alpha_0 & = g n \alpha_0 + g \sqrt{n} \sum_\k \alpha_\k 
    \\ \label{eq:pol2}
    E \alpha_\k & = \ek \alpha_\k + g n \alpha_\k 
    + g \sum_{\k'} \alpha_{\k'} + g \sqrt{n} \alpha_0 + g \sqrt{n} \sum_{\k'} \alpha_{\k\k'}  
    \\ \label{eq:pol3}
    E \alpha_{\k_1\k_2} & = (\epsilon_{\k_1} + \epsilon_{\k_2}) \alpha_{\k_1\k_2} + g \sum_{\k'} \alpha_{\k_1\k'} + g \sum_{\k'} \alpha_{\k'\k_2} + 
    \sum_{\k_1'\k_2'} V(\k_1 - \k_1')\alpha_{\k_1' , \k_1+\k_2-\k_1'}
     + g \sqrt{n} (\alpha_{\k_1} + \alpha_{\k_2}) +
    \ldots 
\end{align}
\end{subequations}
Here we have only kept the leading order boson-boson interaction terms that survive in the dilute limit $n^{1/3}\ab \to 0$. This includes the $V(\q)$ term in Eq.~\eqref{eq:pol3}, which is required to correctly describe the three-body physics.

Let us now rescale everything by $\alpha_0$ such that we have  $\rho_\k = \alpha_\k/\alpha_0$ and $\rho_{\k_1\k_2} = \alpha_{\k_1\k_2}/\alpha_0$. Combining Eqs.~\eqref{eq:pol1} and \eqref{eq:pol2} to replace the energy $E$ then gives
\begin{align}
    0 =  \ek \rho_\k + g \sqrt{n} + \underbrace{g \sum_{\k'} \rho_{\k'}}_{\sqrt{n} \chi} + \underbrace{g \sqrt{n} \left(\sum_{\k'} \rho_{\k\k'} - \rho_\k \sum_{\k'} \rho_{\k'} \right)}_{\rho_\k G_\k} .
\end{align}
Solving for $\rho_\k$ then gives
\begin{align}
    \rho_\k = - \frac{g + \chi}{\ek +G_\k} \sqrt{n} \qquad \textrm{and} \qquad \chi = 
    \left[\frac{1}{g} + \sum_\k \frac{1}{\ek + G_\k} \right]^{-1} \, ,
\end{align}
where we have taken the renormalization condition $g \to 0$ as $\Lambda \to \infty$, and we have assumed that the function $G_\k$ does not diverge as $k \to \infty$.  This finally yields the energy in Eq.~\eqref{eq:energy} of the main text,
\begin{align} \label{eq:energy2}
    E = n \chi = n\left[\frac{m}{2\pi a} + \sum_\k\left(\frac{1}{\ek + G_\k} - \frac{1}{\ek}\right) \right]^{-1} .
\end{align}
The function $G_\k$ can be viewed as the potential energy difference between 1 and 2 boson excitations, thus providing a measure of the quantum correlations between bosons. In the absence of interactions and correlations, we have $G_\k = 0$ and we recover the non-interacting boson limit, as discussed below.
For the case of complete boson blocking where $\rho_{\k\k'} = 0$, the behavior reduces to the Bose polaron version of the Chevy ansatz~\cite{Chevy2006,Li2014}. We then have $G_\k = -E$ according to Eq.~\eqref{eq:pol1} and the definition of $G_\k$.

\subsection{Coherent state ansatz}
In the limit of non-interacting bosons $\ab \to 0$, the many-body wave function is known exactly and corresponds to the coherent state~\cite{Shchadilova2016}:
\begin{align} \label{eq:wfn-coh}
    \ket{\Psi_{\rm coh}} = \mathcal{N} e^{\sum_\k \beta_\k b^\dag_\k} \ket{\Phi} ,
\end{align}
where the normalization factor $\mathcal{N} = e^{-\frac{1}{2}\sum_\k |\beta_\k|^2}$, and $\ket{\Phi}$ is the 
BEC ground state in the absence of the impurity. If we compare this with the general form of the static impurity wave function in Eq.~\eqref{eq:wfn2}, we clearly have $\alpha_0 = \mathcal{N}$, \ $\alpha_\k = \mathcal{N} \beta_\k$, \ $\alpha_{\k_1\k_2} = \mathcal{N} \beta_{\k_1} \beta_{\k_2}$, \ $\alpha_{\k_1\k_2\k_3} = \mathcal{N} \beta_{\k_1} \beta_{\k_2} \beta_{\k_3}$, and so on. Thus, the bosons are completely uncorrelated within the coherent state $\ket{\Psi_{\rm coh}}$.

The Hamiltonian \eqref{eq:Ham-pol} in the limit of an ideal Bose gas is simply
\begin{align}
    \hat{H} = \sum_\k \ek b^\dag_\k b_\k  + g \sqrt{n} \sum_{\k}(b^\dag_\k + b_{-\k}) + g \sum_{\k\q} b^\dag_\k b_\q + g n.
\end{align}
Applying this Hamiltonian to the state $\ket{\Psi_{\rm coh}}$ gives 
\begin{align}
    \hat{H} \ket{\Psi_{\rm coh}} = \left[\sum_\k \left(\ek \beta_\k + g \sqrt{n} + g \sum_\q \beta_\q \right) b^\dag_\k + g \sqrt{n} \sum_\k \beta_\k +   g n \right]   \ket{\Psi_{\rm coh}} ,
\end{align}
where we have used the fact that $b_\k \ket{\Psi_{\rm coh}} = \beta_\k \ket{\Psi_{\rm coh}}$. Therefore, in order for the state $\ket{\Psi_{\rm coh}}$ to be an eigenstate of the Hamiltonian, we require the term involving $b^\dag_\k$ to be zero, which gives the constraint:
\begin{align}
    \ek \beta_\k + g \sqrt{n} + g \sum_\q \beta_\q = 0 .
\end{align}
Solving for $\beta_\k$ and using the renormalization condition then yields the simple expression
\begin{align}
    \beta_\k = - \frac{2\pi a}{m} \frac{\sqrt{n}}{\ek} ,
\end{align}
and the exact eigenenergy just corresponds to the mean-field energy without boson correlations:
\begin{align} \label{eq:energy_coh}
    E_{\rm coh} = g \sqrt{n} \sum_\k \beta_\k + g n  = \frac{2\pi a}{m} n . 
\end{align}
Here we again used the renormalization condition, taking $g\to0$ and $\Lambda\to\infty$.

The energy in Eq.~\eqref{eq:energy_coh} only corresponds to the ground-state energy for $a<0$, while for $a>0$ it corresponds to an excited state that is adiabatically connected to the non-interacting limit $a \to 0$.  
We see that the energy is finite for $a<0$ but it eventually diverges at unitarity $1/a = 0$ when a two-body bound state appears. Moreover, the energy is unbounded for $a>0$, where an arbitrary number of bosons can bind to the impurity. 

Taking boson-boson interactions into account, the coherent-state ansatz is only approximate, with the bare bosonic operators in Eq.~\eqref{eq:wfn-coh} replaced by Bogoliubov quasiparticle operators.
A variational minimization then gives~\cite{Shchadilova2016}
\begin{align}
    \beta_\k = - \frac{2\pi}{m} \frac{1}{a^{-1} - a_0^{-1}}\frac{\sqrt{n \ek}}{E_\k^{3/2}} ,
    \label{eq:betak}
\end{align}
with Bogoliubov dispersion $E_\k = \sqrt{\ek(\ek +8\pi \ab n/m)} \equiv \sqrt{\ek(\ek + 1/m\xi^2)}$, and 
\begin{align}
 a_0^{-1} \equiv  \frac{2\pi}{m} \sum_\k \left(\frac{1}{\ek} -\frac{\ek}{E_\k^2} \right) = \frac{\sqrt{2}}{\xi},
\end{align}
where we have defined the coherence length of the BEC, $\xi=1/\sqrt{8\pi n a_B}$.
Within the coherent-state approximation, the energy is~\cite{Shchadilova2016}
\begin{align}
    E_{\rm coh} = n \left[\frac{m}{2\pi a} + \sum_\k\left(\frac{1}{\ek + 1/m\xi^2} - \frac{1}{\ek}\right) \right]^{-1} 
     =  \frac{2\pi}{m} \frac{n}{a^{-1} - \sqrt{2}/\xi} .
\end{align}
Comparing this to the energy in Eq.~\eqref{eq:energy} of the main text (or Eq.~\eqref{eq:energy2}), we see that the correlation function $G_\k$ is replaced by the constant $1/m\xi^2 \equiv 8\pi \ab n/m$, which is simply the mean-field interaction between an excited boson and the condensate. One can recover this from Eq.~\eqref{eq:pol} by including the lowest-order mean-field terms due to interactions with the condensate and taking the bosons to be uncorrelated, $\alpha_{\k\k'} = \alpha_\k \alpha_{\k'}/\alpha_0$.

\subsubsection{Residue}

In the ideal gas limit, the overlap with the non-interacting state, i.e., the residue, can be calculated within the coherent state ansatz:%
\begin{align}
   Z \equiv |\langle\Phi|\Psi\rangle|^2
    = \exp\left[- \sum_\k \beta^2_\k \right] = \exp\left[- \left(\frac{2\pi a \sqrt{n}}{m}\right)^2 \sum_\k \frac{1}{\ek^2}\right] \to 0 .
\end{align}
This is seen to vanish since the momentum sum is infrared divergent. This divergence is cured once there are interactions between bosons and the coherence length $\xi$ is finite. In this case, using the variational parameters in Eq.~\eqref{eq:betak} and expanding to lowest order in the impurity-boson scattering length $a$ then gives~\cite{Shchadilova2016}
\begin{align} \label{eq:log-residue}
    \ln Z  \simeq -\sum_\k \beta_\k^2 
    & 
    \simeq - \left(\frac{2\pi a}{m}\right)^2 \sum_\k \frac{n}{\sqrt{\ek}(\ek +1/m\xi^2)^{3/2}} \\
     & = - 4\sqrt{2} a^2 \xi n, 
\end{align}
which matches second-order perturbation theory~\cite{Christensen2015}.

On the other hand, we note that the Gross-Pitaevskii approach in Ref.~\cite{Guenther2021} does not yield the correct residue in the weak-coupling limit $n^{1/3} |a| \ll 1$, since it appears to consider everything with respect to the non-interacting BEC rather than the weakly interacting state.  Thus, instead of $-\sum_\k \beta_\k^2$ like in Eq.~\eqref{eq:log-residue}, it gives
\begin{align}
    \ln Z & \simeq 
     - \sum_\k \frac{\ek}{E_\k} \beta_\k^2=- \sqrt{2}\pi a^2 \xi n.
\end{align}

\section{Bosonic Anderson model}

As an alternative manner of introducing correlations between the bosons, we also consider the following model:
\begin{align}
    \hat H=\sum_\k \ek 
    b^\dag_\k 
    b_\k +\nu_0 
    d^\dag 
    d+\lambda\sum_\k(
    d^\dag b_\k+b^\dag_\k d)+\frac U2d^\dag d^\dag dd,
    \label{eq:Ham2}
\end{align}
where we assume that the operator $d$ is bosonic and we take the limit $U\to +\infty$ at the end of the calculation. While this model features no explicit interactions between the bosons, correlations are induced by the presence of the impurity since the impurity-boson interaction changes a boson to the auxilliary state described by the operator $d$ (this models the coupling of an open and a closed interaction channel in a Feshbach resonance~\cite{Chin2010}). The infinite repulsion between $d$ states described by the last term in Eq.~\eqref{eq:Ham2} then ensures that the interaction channel, which depends on the presence of the impurity, is only available to one boson at a time. Due to the similarity with the Anderson impurity model, we refer to Eq.~\eqref{eq:Ham2} as the ``bosonic Anderson model''. The few-body physics of this model was investigated in detail in Ref.~\cite{Shi2018,Yoshida2018PRA}.

\subsection{Truncated basis approach} 

To investigate the ground state of a Bose polaron within the bosonic Anderson model, we apply a variational principle using the truncated basis ansatz.  In this approach, we approximate the polaron ground state as
\begin{align}
\ket{\Psi}=\ket{\psi_0}+\ket{\psi_1}+\ket{\psi_2}+\ket{\psi_3},
\end{align}
where the states with different numbers of excitations are:
\begin{align}
\ket{\psi_0} = & 
    \alpha_0 \ket{\Phi}\nn \\
\ket{\psi_1} = & 
\left(\sum_{\k\neq0} \alpha_\k b^\dag_\k
 + \gamma_0 d^\dag  \right) \ket{\Phi}\nn \\
\ket{\psi_2} = & 
\left(\frac{1}{2} \sum_{\k_1, \k_2 \neq 0} \alpha_{\k_1 \k_2}  b^\dag_{\k_1} b^\dag_{\k_2} 
    + \sum_{\k\neq 0} \gamma_\k d^\dag b^\dag_\k 
\right) \ket{\Phi}\nn \\
\ket{\psi_3} = &   \Bigg(\frac{1}{6} \sum_{\k_1, \k_2, \k_3 \neq 0} \alpha_{\k_1 \k_2 \k_3}
         b^\dag_{\k_1} b^\dag_{\k_2}
                 b^\dag_{\k_3} + \frac{1}{2}\sum_{\k_1, \k_2 \neq 0} \gamma_{\k_1 \k_2} d^\dag 
        b^\dag_{\k_1} b^\dag_{\k_2}
                  \Bigg) \ket{\Phi}.
\label{eq:psi}
\end{align}
We do not include states that have multiple bosons in the $d$ state because such states are prohibited by the limit $U\to\infty$, as explained above.
The variational parameters are determined by the variational equation $\partial_{\alpha^\ast,\gamma^\ast}\matrixel{\Psi}{(\hat H - E)}{\Psi}=0$, which reads
\begin{subequations}
\begin{align}
    E \alpha_0 &= \lambda \sqrt{n} \gamma_0, \\
    E \alpha_{\k}
        &= \ek \alpha_\k
        + \lambda \gamma_0
        + \lambda \sqrt{n} \gamma_\k, \\
    E \alpha_{\k_1 \k_2}
        &= (\epsilon_{\k_1} + \epsilon_{\k_2}) \alpha_{\k_1 \k_2}
        + \lambda (\gamma_{\k_1} + \gamma_{\k_2})
        + \lambda \sqrt{n} \gamma_{\k_1 \k_2}, \\
    E \alpha_{\k_1 \k_2 \k_3}
        &= (\epsilon_{\k_1} + \epsilon_{\k_2} + \epsilon_{\k_3}) \alpha_{\k_1 \k_2 \k_3}
        + \lambda (\gamma_{\k_1 \k_2} + \gamma_{\k_2 \k_3} + \gamma_{\k_3 \k_1}), \\
    E \gamma_0
        &= \nu_0 \gamma_0
        + \lambda \sqrt{n} \alpha_0
        + \lambda \sum_{\k\neq 0} \alpha_\k, \\
    E \gamma_\k
        &= (\nu_0 + \epsilon_\k) \gamma_\k
        + \lambda \sqrt{n} \alpha_\k
        + \lambda \sum_{\k'} \alpha_{\k \k'}, \\
    E \gamma_{\k_1 \k_2}
        &= (\nu_0 + \epsilon_{\k_1} + \epsilon_{\k_2}) \gamma_{\k_1 \k_2}
        + \lambda \sqrt{n} \alpha_{\k_1 \k_2}
        + \lambda \sum_{\k'} \alpha_{\k_1 \k_2 \k'}.
\end{align}
\end{subequations}

These coupled equations are ultraviolet divergent with fixed bare parameters $\nu_0$ and $\lambda$.
To remedy this, we rewrite them using physical low-energy parameters.
To this end, we first remove the $\alpha$ coefficients and obtain
\begin{subequations}
\begin{align}
    \mathcal{T}_A^{-1}(E) \gamma_0
        &= \frac{n \gamma_0}{E}
        + \sum_{\k} \frac{\sqrt{n} \gamma_\k}{E-\ek}, \label{eq:var1}\\
    \mathcal{T}_A^{-1}(E-\ek) \gamma_\k
        &= \frac{\sqrt{n} \gamma_0}{E-\ek}
        + \frac{n\gamma_\k}{E-\ek}
        + \sum_{\k'} \left(
            \frac{\gamma_{\k'}}{E-\ek-\epsilon_{\k'}}
            +  \frac{\sqrt{n}\gamma_{\k \k'}}{E - \epsilon_{\k} - \epsilon_{\k'}}
        \right), \label{eq:var2} \\
    \mathcal{T}_A^{-1}(E-\epsilon_{\k_1}-\epsilon_{\k_2}) \gamma_{\k_1 \k_2}
        &= \frac{\sqrt{n} (\gamma_{\k_1} + \gamma_{\k_2})}{E - \epsilon_{\k_1} - \epsilon_{\k_2}}
        + \frac{n\gamma_{\k_1 \k_2}}{E-\epsilon_{\k_1}-\epsilon_{\k_2}}
        + \sum_{\k'} \frac{\gamma_{\k_1 \k'} + \gamma_{\k_2 \k'}}{E - \epsilon_{\k_1} - \epsilon_{\k_2} - \epsilon_{\k'}}. \label{eq:var3}
\end{align}
\end{subequations}
Here, we have defined the scattering $T$ matrix in the Anderson model
\begin{align}
    \mathcal{T}_A^{-1}(E)
        = \frac{1}{\lambda^2} (E-\nu_0)
        - \sum_\k \frac{1}{E - \ek}.
\end{align}
To eliminate the ultraviolet divergence of the momentum sum, we carry out the renormalization procedure with a momentum cutoff $\Lambda$ and obtain
\begin{align}
    \mathcal{T}_A^{-1}(E)
        = \frac{m}{2\pi} (
            1/a - r_\mathrm{eff} mE - \sqrt{-2mE -i0}
        ),
    \label{eq:TAnderson}
\end{align}
where $a \equiv (\frac{2\Lambda\nu_0}{\pi}-\frac{2\pi}{m\lambda^2})^{-1}$ is the $s$-wave scattering length, and $r_\mathrm{eff}\equiv - \frac{2\pi}{m^2 \lambda^2}$ is the effective range.
Note that this is identical to the vacuum impurity-boson scattering $T$ matrix in the two-channel model~\cite{Timmermanns1999}.
The resulting equations~(\ref{eq:var1}--\ref{eq:var3}), with $T^{-1}_A$ given in Eq.~\eqref{eq:TAnderson}, do not contain the bare parameters and are ultraviolet finite, and we solved them by discretization.

\begin{figure}
	\centering
	\includegraphics[width=.5\linewidth]{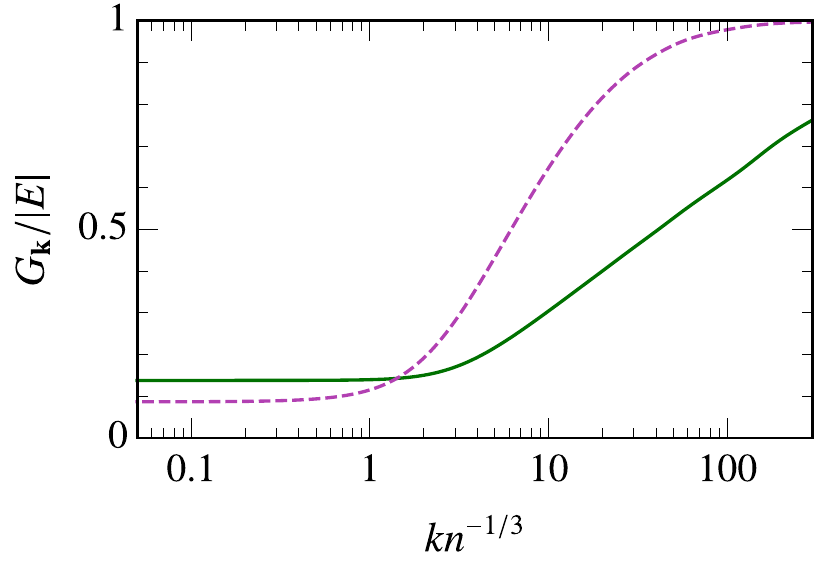}
	\caption{The function $G_\k$ that quantifies boson-boson correlations. We show the results calculated within the bosonic Anderson model using a variational state with three excitations. The lines correspond to those of Fig.~\ref{fig:manybody} of the main text, namely we have $1/a=0$ and $n^{1/3}a^*=0.215$ (purple dashed) or $n^{1/3}a^*=0.00215$ (green solid).}
	\label{fig:Gfunc}
\end{figure}

In Fig.~\ref{fig:Gfunc}, we illustrate the quantum correlations captured by the truncated basis approach, as encoded in the function $G_\k$ from Eq.~\eqref{eq:Gk} of the main text. We focus on the unitarity limit $1/a=0$ where the behavior only depends on the three-body parameter $n^{1/3}a^*$, with $a^*=3.1426|r_{\rm eff}|$~\cite{Shi2018,Yoshida2018PRA}. We see that $G_\k \to -E$ as $k \to \infty$, which corresponds to boson blocking at short distances. Our truncated basis approach also captures how the blocking effect decreases at larger distances (smaller $k$), but it tends to overestimate the repulsion in the infrared limit $k \to 0$, and this error increases with decreasing $n^{1/3}a^*$ for a fixed number of excitations. 
The exact function $G_\k$ in this model is expected to go to zero when $k \to 0$ since it should recover the behavior of uncorrelated bosons with $\ab = 0$.

\end{document}